\newcommand\Sul{Sul--Cu$_2$Cl$_4$}

\documentclass[prb,twocolumn,floatfix,preprintnumbers,amsmath,amssymb]{revtex4}

\usepackage{graphicx}
\usepackage{bm}

\begin{document}

\title{Dimensional crossover in a spin liquid to helimagnet quantum phase transition}

\author{V. O. Garlea}
 \email{garleao@ornl.gov}
\affiliation{Neutron Scattering Sciences Division, Oak Ridge National Laboratory, Oak Ridge, Tennessee 37831, USA.}
\author{A. Zheludev}
\affiliation{Neutron Scattering Sciences Division, Oak Ridge National Laboratory, Oak Ridge,
Tennessee 37831, USA.}
\author{K. Habicht}
\author{M. Meissner}
\affiliation{BENSC, Hahn-Meitner Institut, D-14109 Berlin, Germany.}
\author{B.~Grenier}
\author{L.-P.~Regnault}
\author{E.~Ressouche}
\affiliation{CEA-Grenoble, INAC-SPSMS-MDN, 17 rue des Martyrs,
38054 Grenoble Cedex 9, France.}
\date{\today}

\begin{abstract}
Neutron scattering is used to study magnetic field induced
ordering in the quasi-1D quantum spin-tube compound \Sul\ that in
zero field has a non-magnetic spin-liquid ground state. The
experiments reveal an incommensurate chiral high-field phase
stabilized by a geometric frustration of the magnetic
interactions. The measured critical exponents $\beta\approx0.235$ and
$\nu\approx0.34$ at $H_c\approx3.7$~T point to an unusual sub-critical
scaling regime and may reflect the chiral nature of the quantum
critical point.
\end{abstract}


\maketitle

Quantum critical points (QCPs) in spin liquids have recently
become a forefront issue in magnetism.~\cite{Sachdev08} In
particular, phase transitions in gapped quantum-disordered
antiferromagnets (AFs) induced by the application of external
magnetic fields provide a new way to study a number of fundamental
phenomena. For example, some models can be directly mapped onto
Bose-Einstein condensation (BEC),~\cite{Rice02,Giamarchi99,Giamarchi08} while fractional
magnetization plateaus  in certain systems are magnetic analogues
of Mott-insulator phases,~\cite{Oshikawa97,Rice02} and quenched
disorder can lead to the formation of an effective Bose glass.~\cite{Nohadini05} At the same time, the abundance and variety of
low-dimensional spin systems enable access to an entire range of
previously inaccessible dimensional-crossover phenomena.~\cite{Watson01,Lake05,Sebastian06,Batista07,Orignac07} Most
importantly, field-induced quantum phase transitions are found in
a number of prototypical magnetic materials, and can thus be
studied experimentally. Neutron scattering turned out to be particularly useful, driving much of
the theoretical development.~\cite{Chen01,Ruegg03,Grenier04,Garlea07}

Another current topic in quantum magnetism is that of chirality.
Spontaneous breaking of inversion symmetry in classical magnets
has been known for many decades and manifests itself in long-range
helimagnetic order. Today, theorists seek to understand how
chirality can exist in disordered spin liquids,~\cite{Lecheminat05,Yao07,Sato07} and how it may be involved in
quantum phase transitions and critical behavior.~\cite{Momoi03,Hasenbusch05} In this context, we hereby report an
experimental observation of a field-induced quantum phase
transition from a disordered spin liquid to an ordered chiral
incommensurate state. This phenomenon is studied in the
geometrically frustrated Heisenberg S=1/2 spin-tube
antiferromagnet~\Sul. We observe highly unusual values of the
critical exponents that represent dimensional crossover scaling
and may be a signature of chirality at this QCP.

As discussed in detail in Ref~\onlinecite{GarleaSUL}, \Sul, with the
chemical formula Cu$_2$Cl$_4$-H$_8$C$_4$SO$_2$, realizes a rare
$S=1/2$ four-leg Heisenberg spin tube model, the tubes running
along the $c$ axis of the triclinic crystal structure.~\cite{GarleaSUL}
Zero-point quantum spin fluctuations entirely destroy long-range order in this system.
The magnetic ground state is a spin liquid, with activated susceptibility and
specific heat. The spectrum consists of strongly dispersive
triplet excitations with an energy gap of $\Delta\approx 0.52$~meV
and a spin wave velocity of $v\approx14$~meV. Neutron scattering
experiments failed to detect any dispersion of magnetic
excitations perpendicular to the tube axis or any splitting of the
gap mode in zero field. Based on the experimental resolution of
about 0.2~meV FWHM, one can place upper bounds on the inter-tube
coupling and anisotropic (non-Heisenberg) interactions: $J_\bot, D
<0.05$~meV, respectively. Thus, \Sul\ is an exceptionally
isotropic and 1-dimensional system. The conveniently small gap can
be overcome by applying a moderate magnetic
field.~\cite{Fujisawa03,Fujisawa05} A field-induced ordering
transition occurs at $H_c\approx 4$~T and manifests itself in a
lambda specific-heat anomaly and the appearance of non-zero
uniform magnetization.~\cite{Fujisawa03,Fujisawa05} The
directional dependence of the critical field is fully accounted
for by the anisotropy of the $g$-tensor.~\cite{Fujisawa03} This
gives an even tighter limit on the magnitude of anisotropy:
$D<10^{-3}$~meV.

A crucial feature of \Sul\ is a partial geometric frustration of
exchange interactions on the tube rungs.~\cite{GarleaSUL} In a
classical magnet such frustration is often resolved through the
formation of a spiral (helimagnetic) spin
structure.~\cite{Yoshimori} The latter has a periodicity defined by
the ratio of conflicting exchange constants and therefore totally
independent of the periodicity of the underlying crystal lattice.
Due to the singlet nature of the ground state in \Sul, such static
helimagnetic order is absent, but {\it dynamic} incommensurate
correlations are preserved. The equal-time correlation function is
maximized, and the gap modes have dispersion minima at
incommensurate positions, slightly off the AF point:
$l_0=0.5\pm\delta$, $\delta=0.022(2)$.~\cite{GarleaSUL}

The field-induced QCP was studied in two series of neutron
scattering experiments. On the V2-FLEX 3-axis spectrometer at HMI
we utilized an assembly of 12 fully deuterated \Sul\ single
crystals with a total mass of about 1~g. The crystals were
co-aligned to a triangular mosaic spread of  1.9$^{\circ}$ full
width at half maximum (FWHM). The $(h,0,l)$ reciprocal-space plane
coincided with the scattering plane of the spectrometer, while the
magnetic field, generated by a 14.5~T cryomagnet, was applied
along the $b$ axis. The instrument was operated in 3-axis mode,
using 4.7~\AA\ neutrons selected by and Pyrolitic Graphite PG(002)
monochromator and analyzer. Sample temperature was controlled by a
$^3$He-$^4$He dilution refrigerator. A second series of
measurements was carried out on the D23 lifting-counter
diffractometer at ILL. Diffraction data were taken on a 2.5 x 2 x
5 mm$^3$ deuterated single crystal, whose quality was previously
tested using the Orient Express instrument at ILL. The crystal was
loaded in a 12~T superconducting magnet ($H\parallel b$) equipped
with a dilution insert. Diffracted peaks in the $(h, 0, l)$ and
$(h, 1, l)$ scattering planes were measured using monochromatic
neutron beams with the wavelengths 1.27~$\AA$ (Cu 200) or
2.37~$\AA$ (PG 002).

\begin{figure}[tp]
\includegraphics[width=3.55in]{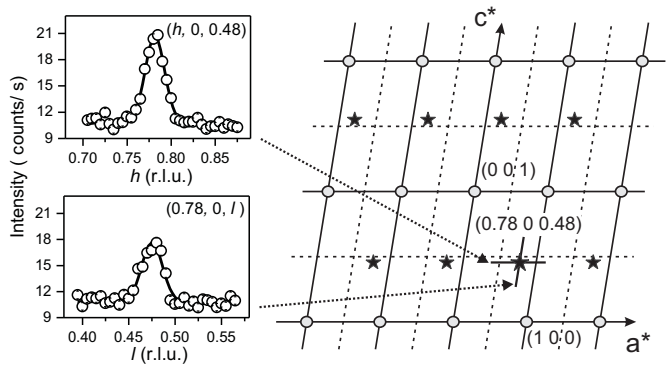}
\caption{Neutron scattering elastic scans along $a^{*}$ and
$c^{*}$ directions through the $(0.78, 0, 0.48)$ magnetic peak,
and reciprocal lattice space of \Sul indicating the incommensurate
nature of the field induced long-range magnetic order.}
\label{reciplatt}
\end{figure}

Neutron data collected at $T=130$~mK confirm that the
field-induced transition previously seen in bulk measurements
represents the onset of long-range magnetic order. A comprehensive
search in the $(h, 0, l)$ plane detected the emergence of new
magnetic Bragg peaks beyond $H_\mathrm{c}=3.7$~T. The value of the
critical field is lower than the value 4.3~T reported in
Ref.~\onlinecite{Fujisawa03}. The discrepancy is accounted for by a
lower experimental temperature, a different geometry ($H\parallel b$) and the anisotropy of the $g$-tensor.~\cite{Fujisawa05}
The observed magnetic reflections are located at incommensurate reciprocal-space
positions. As shown in Fig.~\ref{reciplatt}, these peaks can
be indexed as $(h\mp\xi,k,l\pm\zeta)$, where $\xi$=0.22,
$\zeta$=0.48 and $h$, $k$ and $l$ are integers. Peaks
corresponding to odd values of $l$ are systematically absent.
Further searches took advantage of the lifting counter geometry of
D23 venturing outside the $(h,0,l)$ plane but revealed no
additional reflections.

\begin{figure}[tp]
\includegraphics[width=3.55in]{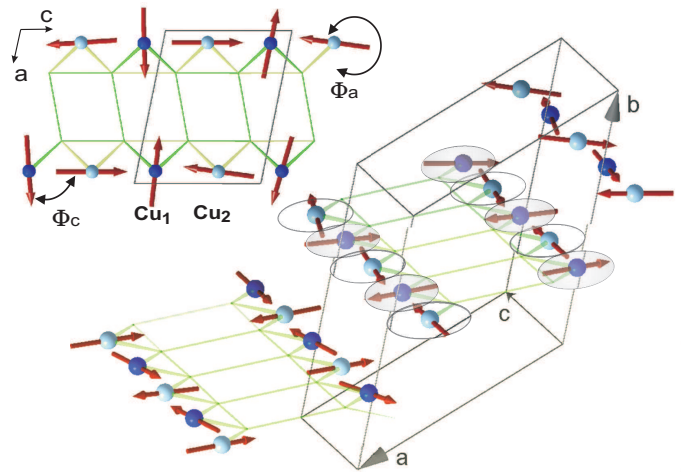}
\caption{(Color online) Schematic view of the incommensurate magnetic phase,
modeled by a planar helimagnetic structure with the wave vector
$k=(-0.22, 0, 0.48)$.}
\label{magstr}
\end{figure}

The structure of the high-field ordered phase was determined from
23 independent magnetic Bragg intensities measured at $T=60$~mK in
a field $H=10$~T applied along the $b$ direction. The obtained
data set is not sufficient for an unambiguous model-independent
determination of the spin arrangement in the incommensurate case.
Nevertheless, a unique solution can be derived under just a few
additional assumptions. As mentioned above, in classical
Heisenberg magnets geometric frustration typically favors a planar
helimagnetic state.~\cite{Yoshimori} Such spin configurations
survive in quantum spin models,~\cite{White96} albeit with
strongly renormalized periodicities. In our data analysis we
therefore postulated a uniform helimagnetic structure for \Sul\ as
well. As shown in Fig.~\ref{magstr}, all spins were assumed to be
confined in the plane perpendicular to the direction of applied
field and the period of the spin-spiral was chosen to match the
observed magnetic propagation vector. The basis of this helical
structure is defined by two relative pitch angles $\phi_a$ and
$\phi_c$ between spins coupled along the $a$ and $c$ axes in each
unit cell, respectively (Fig.~\ref{magstr}). A least-squares fit
of this model to the data yields an excellent agreement with
$\phi_a= 273\pm3^\circ$, $\phi_c=83\pm9^\circ$ and an ordered
moment $m=0.044(1)~\mu_\mathrm{B}$ per site. The obtained solution
is unique within the assumed planar-spiral model with 4 spins per
unit cell.

\begin{figure}[bp!]
\includegraphics[width=3.3in]{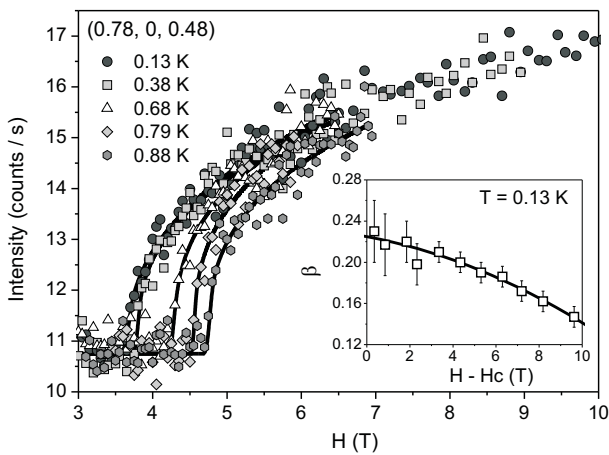}
\caption{Field dependence of the $(0.78, 0, 0.48)$ peak intensity measured at  selected temperatures (symbols). The solid lines
are power-law fits to the data as described in the text. Inset: The value of the critical index $\beta$ resulted from fitting
the data over a progressively smaller field range ($H-H_{c}$) at T = 0.13 K. The line is a guide for the
eye.}\label{orderparam}
\end{figure}

\begin{figure} \includegraphics[width=3.3in]{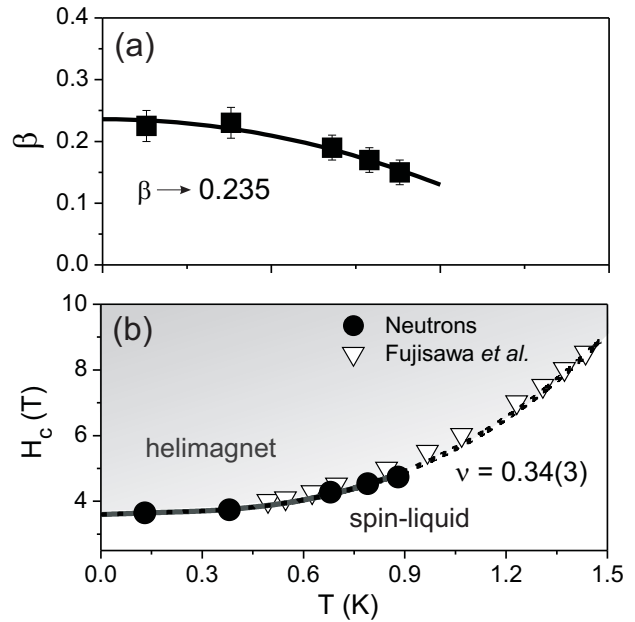}
\caption{(a) The order parameter critical index $\beta$ plotted as a function of temperature. The T = 0~K extrapolated value, $\beta\rightarrow0.235$, is in clear disagreement with that expected for a 3D BEC. (b) Temperature dependence of the
transition field, $H_\mathrm{c}$ determined by neutron scattering (solid circles) and specific heat study (open triangles,
Ref.~\onlinecite{Fujisawa05}). Solid line represents a power-law fit to the neutron data, yielding a critical index $\nu=0.34(3)$}
\label{phdiag}
\end{figure}

To access the character of the phase transition itself, we
measured the critical exponent $\beta$ associated with the magnetic
order parameter $m(H,T)$ and defined as
$m(H,T)\propto(H-H_\mathrm{c}(T))^{\beta(T)}$ for $H \rightarrow
H_\mathrm{c}$. The field dependencies of the $(0.78, 0, 0.48)$
peak intensity, expected to be proportional to $|m|^2$, was
measured at several temperatures and is plotted in Fig.~\ref{orderparam}. As exemplified in the inset of
Fig.~\ref{orderparam} for the case of $T=130$~mK, power-law fits
to the data (Fig.~\ref{orderparam}, solid lines) were performed
over a progressively shrinking field range $\delta H$. Taking the
limit $\delta H \rightarrow 0$ at each temperature allows us to
zero in on the actual critical region. The resulting $\beta(T)$
and $H_\mathrm{c}(T)$ are plotted in solid symbols in
Fig.~\ref{phdiag}. The typical error bar on $H_\mathrm{c}$ is
0.02~T. Temperature dependence of the exponent $\beta$ was  empirically fit to a
parabola that had zero slope at T = 0~K. The zero-temperature extrapolation of the parabola yielded a value $\beta=0.235(6)$.
Another important critical index is $\nu$ that defines the phase boundary:
$H_\mathrm{c}(T)-H_\mathrm{c}(0)\propto T^{1/\nu}$ at
$T\rightarrow 0$. From a power-law fit (solid line in
Fig.~\ref{phdiag}b) to our $H_\mathrm{c}(T)$ data we get
$\nu=0.34(3)$. Overall, the fitting curve agrees well with the
results of bulk measurements~\cite{Fujisawa05} (open symbols in
Fig.~\ref{phdiag}b).

The measured value of the critical exponents are quite unusual. In
particular, they are obviously different from those in the
well-understood scenario of 3D BEC of magnons,~\cite{Giamarchi99,Giamarchi08} where $\beta=0.5$ and
$\nu=2/d=2/3$. The distinction is not entirely unexpected, as a
description of the excitations in terms of dilute hardcore bosons
may no longer hold in the incommensurate case. In the following
paragraphs we shall separately consider three possible reasons for
the exotic quantum critical scaling.

First, we can totally rule out the effect of non-Heisenberg terms
in the spin Hamiltonian. These play a key role in defining the
character of the phase transition in some other spin-gap
materials, such as the $S=1$ spin chain compound Ni(C$_5$D$_{14}$N$_{2}$)$_{2}$N$_{3}$(PF$_{6}$)
(NDMAP).~\cite{Chen01} However, in our case of \Sul, the smallest
field interval used in the determination of $\beta$ and
$H_\mathrm{c}$ is $\delta H_\mathrm{min}=0.5$~T. This window
defines the energy scale of the slowest relevant fluctuations
$\hbar\omega_\mathrm{min}=g \mu_B \delta H_\mathrm{min}\sim
0.1$~meV. Since $D\ll \hbar\omega_\mathrm{min}$, any anisotropy
terms in the Hamiltonian will manifest themselves only much closer
to the critical point than our analysis can approach.

Much more relevant is the question of whether our experiments can
access the true critical indexes of 3D long-range ordering in a
material as effectively 1D as \Sul. In fact, by the same reasoning
as in the previous paragraph, they can {\it not}, as $J_\bot<
\hbar\omega_\mathrm{min}$. The 3D critical indexes manifests themselves only
undetectably, close to the transition point. In the absence of residual 3D coupling, the
quasi-1D \Sul\ would remain disordered at $T=0$ even in an strong applied fields. Instead, it would become a Luttinger spin liquid,~\cite{Sachdev} with a divergent correlation length but no static long-range order.

Dimensional crossover at the field-induced QCP in the relevant
quasi-1D case was recently studied in the context of the NMR spin
relaxtion rate $1/T_1$ in the {\it disordered} state.~\cite{Orignac07} Though the critical exponents associated with
the {\it ordered} phase and measured in this work have not yet
been investigated theoretically, one can draw some analogies with
the thermodynamic phase transition in classical quasi-2D XY
magnets. As a function of temperature, the 2D system does not
order in the usual sense, though the correlation length diverges
at the Kosterlitz-Thouless point.~\cite{KT} The experimentally
observable 3D ordering in layered materials is governed by a
universal ``sub-critical'' exponent $\beta=0.23$,~\cite{Bramwell}
distinct from the true 3D-XY critical index $\beta=0.35$.~\cite{Campostrini} The scaling observed
in \Sul\ will correspond to an analogous sub-critical regime, but
whether or not the exponents are universal is yet to be established.

The third and most intriguing consideration is that the
(sub)critical indexes in \Sul\ are modified by the chiral nature
of the ordered state. As famously conjectured by
Kawamura,~\cite{Kawamura} helimagnetic ordering forms separate
chiral universality classes with distinct critical indexes. Though
still controversial, this theory has been apparently confirmed
experimentally in a number of frustrated triangular-lattice AFs,~\cite{Mason} and may apply to the QCP in \Sul. Even more
interesting is the possibility that due to strong geometric
frustration chirality is already present in the spin liquid phase
of \Sul. The existence of such chiral spin liquids is now well
established for spin ladders with 4-spin exchange,~\cite{Lecheminat05} as well as for the Kitaev model.~\cite{Yao07}
Thus one can imagine a scenario where chirality in \Sul\ is
present in zero field or appears  in a separate phase transition
at $H<H_\mathrm{c}$.

In summary, we have observed a field-induced QCP that separates a
gapped spin liquid state from an incommensurate chiral
helimagnetic state in a quasi-1D frustrated quantum AF. The highly
unusual values of the order-parameter critical exponents pose
three important questions to be answered by theorists. Is this QCP
characterized by an extended subcritical scaling regime in the
{\it ordered} state and is this scaling universal? Does the
chirality of the ordered state qualitatively alter the critical
and/or sub-critical behavior? Or, does chirality appear at lower
fields, before the onset of long range order?

The authors thank M. Boehm for the help provided during preliminary measurements and R. Custelcean (ORNL) for his input into the crystal structure analysis. A meaningful discussion of the results would be impossible without the intellectual guidance
provided by F. Essler, O. Tchernyshev and I. Zaliznyak. Research at ORNL was funded by the United States Department of Energy,
Office of Basic Energy Sciences- Materials Science, under Contract No. DE-AC05-00OR22725 with UT-Battelle, LLC.

\end{document}